Z.Zh. Zhanabaev, A.T. Agishev, S.A. Khokhlov

# NONLINEAR FRACTAL MEANING OF THE HUBBLE CONSTANT

*Al-Farabi Kazakh National University,*
*Al-Farabi Avenue, 71, Almaty, 050040, Kazakhstan*
*\*E-mail: aldiyar.agishev@gmail.com*

**According to astrophysical observations value of recession velocity in a certain point is proportional to a distance to this point. The proportionality coefficient is the Hubble constant measured with 5% accuracy. It is used in many cosmological theories describing dark energy, dark matter, baryons, and their relation with the cosmological constant introduced by Einstein.**

**In the present work we have determined a limit value of the global Hubble constant (in a big distance from a point of observations) theoretically without using any empirical constants on the base of our own fractal model used for the description a relation between distance to an observed galaxy and coordinate of its center. The distance has been defined as a nonlinear fractal measure with scale of measurement corresponding to a deviation of the measure from its fixed value (zero-gravity radius). We have suggested a model of specific anisotropic fractal for simulation a radial Universe expansion. Our theoretical results have shown existence of an inverse proportionality between accuracy of determination the Hubble constant and accuracy of calculation a coordinates of galaxies leading to ambiguity results obtained at cosmological observations.**

In relativistic cosmology, time dependence of space curvature can be determined via the Einstein field equations describing gravitational interaction. The well-known models describing different space curvatures are zero (Euclidean) model, positive (closed) model, and negative (open) model. All these models describe the singularity known as the Big Bang. For the description of the accelerated expansion of the Universe we can add the cosmological constant $\Lambda$ to the Einstein equations. Comprehension of meaning of this constant requires using of new concepts which are not related to either matter or gravitational field. Using the constant $\Lambda$ which physical meaning isn't completely clear indicates to a necessity of using another approach to solve the problem. Thus, it was stated that "at present there are no persistent and convincing grounds, both observational and theoretical, for such a modification of the basic equations of theory" [1].

The best correspondence between theoretical results and results of observations has been presented in [2,3]. Description of topology of the Universe as a dodecahedral Poincaré space has been suggested in [2]. According to this paper, anisotropy of energy density of cosmic background can be described due to finiteness of spectrum width characterizing this space. The measured value of the Hubble constant has been used as a base of the study. The diagram describing a dependence of velocity on distance to a galaxy is based on data received from «Hubble space telescope» constructed via a nonlinear theory including force of "universal anti-gravity" proportional to the cosmological constant [3].

Despite a huge number of studies, the main facts of observational cosmology such as singularities of galaxy clustering, chaotic arrangement of galaxies near zero-gravity areas, accelerated recession of galaxies, existence of the global Hubble constant, etc., can't be described logically without using unreasonable constants [4, 5]. Authors of the works [2-6] have noted a necessity of construction a space-time theory describing space topology in accordance with physical processes occurring in this space.

So, the aim of our work is to construct a geometrical model for the description the Universe expansion without using any constants with unclear physical meaning, and define the asymptotical (global) value of the Hubble constant. Statement of the problem is illustrated in Fig. 1.

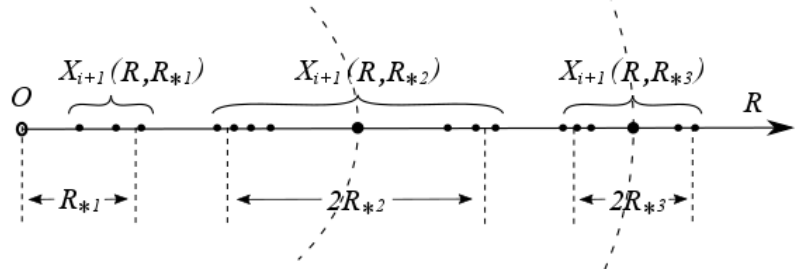

Fig. 1 – Arrangement of galaxies in an arbitrary direction near zero-gravity points $R_*$ on the distance $R$ from an observation point $O$. Location of a galaxy $X_{i+1}(R)$ is described by Eq. (2), order of iteration is i + 1, and $X_{i+1}(R) > R$. Although a point $R_{*3} < R_{*2}$ is located farther than $R_{*2}$, its effect on $X_{i+1}(R)$ is small because intermittent jets of $X_{i+1}(R)$ possible at $R \sim R_*$. So, we use $R_{*j}$, where $j = 1, 2, ...$ for constructing a global model of Universe expansion.

Let`s consider distribution of Galaxies in a homogeneous isotropic space with taking into account gravitational waves. Existence of these waves follows from the Einstein's equations for gravitational fields. In a co-moving frame of reference a time dependence of coordinate $R(t)$ of a galaxy in any direction can be defined with accuracy $\Delta R(t)$ because of wave structure of the space. Therefore, we can consider value of speed in a certain direction as a non-zero constant expressed as $V_0 = \pm \frac{dR(t)}{dt}$.

Speed of a galaxy with coordinate $R$ in the selected system is equal to zero by definition, but if location of the galaxy is chaotically different from $R$, we can describe the velocity as a nonzero value measured in units $V_0$. Determination of velocity of galaxies in the conventional units $V_0$ allows us to replace the time units of measurement of the Hubble constant $H$ by a spatial one which is convenient for separation geometric and space-time types of analysis. As usual [1], $H$ is defined in $[km \cdot s^{-1} Mpc^{-1}]$, therefore, we can accept $V_0 = 1 \, km/s$.

Fluctuations of matter and gravitational interaction of galaxies lead to appearance of local heterogeneities, for example, twisting (clustering) of galaxies. Let us designate distance between a barycenter of a cluster to its boundary (zero-gravity radius) as $R_*$. Distribution of galaxies around $R_*$ is obviously uneven and structural due to the already mentioned conditions which are fluctuations of cosmic matter and heterogeneity of gravitational effects. We take into account the fact that small and large-scale clusters have a similar structure. Objects with structural and scale-invariant structure can be considered as fractals [7]. Since the B. Mandelbrot works, fractal theory is used to describe structure of the Universe on a large scale [8]. However, for more detailed and appropriate description of cosmological observations, we need new concepts of fractal geometry.

According to well-known fractal theories, choice of measurement scale of a measure (we consider the measure as an additive geometric or physical value) should be independent on the measure. Near critical values of parameters, for example, at $R \approx R_*$ the measure varies greatly. So, this singularity should be taken into account. According to the Hausdorff formula, fractal length $X(R)$ can be defined as:

$$X(R) = R \left(\frac{\Delta X}{R_*}\right)^d \cdot \left(\frac{\Delta X}{R_*}\right)^{-D} = R \left(\left|1 - \frac{X}{R_*}\right|\right)^{-\gamma}, \Delta X = |X - R_*|, \gamma = D - d, \quad (1)$$

where $d$ and $D$ are topological and fractal dimensions of $X(R)$, correspondingly.

Eq. (1) should be analyzed with taking into account fluctuations of cosmic matter. So, let us transform Eq.(1) to the following discrete form for spatial evolution [9] written as:

$$X_{i+1}(R) = R \left(\left|1 - \frac{X_i}{R_*}\right|\right)^{-\gamma}. \quad (2)$$

Here $i$ is the spatial step, $i + 1$ in the continuous function $X_{i+1}(R)$ corresponds to the steady-state mode described by Eq.(2) independently on initial conditions. In the cosmological equation

[1], $X_{i+1}(R)$ is a possible value of distance to a galaxy with coordinate $R$. The right side of Eq.(2) describes the multiplication of $R$ to the scale factor.

Values $X_{i+1}(R)$ become chaotic with increasing of parameters $\gamma$ and $R_*$ (Fig. 2a). Rewriting Eq.(2) as a system $X_{i+1} = f(Y_i)$, $Y_{i+1} = f(Z_i)$, $Z_{i+1} = f(X_i)$ we have a possibility to analyze expansion of the Universe (Fig. 2b).

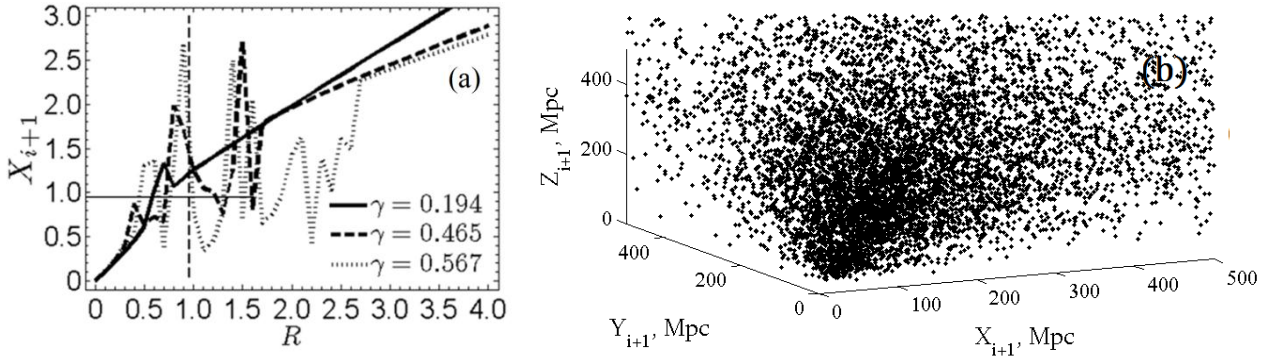

Fig. 2 a) – Chaotization of $X_{i+1}(R)$ with increasing of scaling factor ($\gamma$) at $R_* = 0.96$, number of iterations used for solving Eq.(2) is $10^3$; b) – chaotic map describing distribution of galaxies according to the system of Eq.(2) at $\gamma = 0,4649$, $R_* = 50 \div 250$.

We are interested in the description of asymptotic behavior of $X_{i+1}(R)$ for the set of $R_*$. Such behavior is typical in global cosmology. Therefore, for calculation of final value of the Hubble constant $H$ we don't use $R_*$. Let us describe our choice of $\gamma$. The problem is to search an algorithm for fractalization of an element with length $\Delta R$ (accuracy is $R$) characterized by its maximum of $X_{i+1}(R)$ at $R = R_*$. We must take into account $\Delta R$ because this value is related with $\Delta H$ (accuracy of determination of the Hubble constant). Indeed, existence of gravitational waves follows from the Einstein equations taking into account weak perturbations of the Galilean metric space. For determination regularities of a wave process with characteristic length $\lambda$ and wavenumber k we consider spatial resolution in the minimal interval (from half of wavelength to wavelength) as:

$$\Delta R = \lambda, \ \Delta H = k = 2\pi/\lambda, \ \pi \leq \Delta R \cdot \Delta H \leq 2\pi. \tag{3}$$

Accuracy of a direct measurement is greater than accuracy of an indirect measurement ($\Delta R < \Delta H$). This fact should be taken into account [10] by use of an additional condition.

The anisotropic fractal shown in Fig. 3 has been suggested by one of the authors of this paper [11] to describe turbulence in a boundary layer.

Near zero-gravity ($R \approx R_*$) values $X_{i+1}(R)$ in radial direction reach their maximal values at the growth of the pre-fractal number (order of fractal hierarchy).

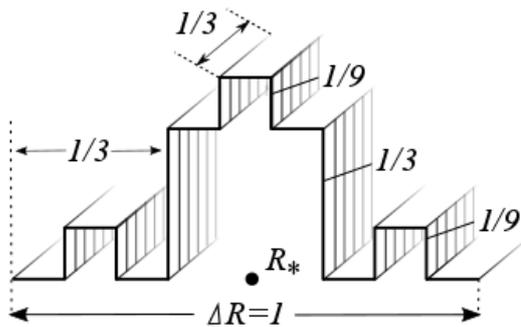

Fig. 3 – 2nd generation ($n = 2$) of the anisotropic fractal, lines of the fractal in the selected (vertical) direction are not deformed. In isotropic space with transversal gravitational waves in the intervals $R_* \pm \Delta R/2$, we have $X_{i+1}(R) > R$ along $R$ direction. Values of $X_{i+1}(R)$ are dependent on height of the $n$-prefractal elements. We accepted $\Delta R = 1$. In general, $\Delta R$ can be determined via the condition for accuracy of definition the Hubble constant according to the Eq.(3).

The Hausdorff local dimension of the anisotropic fractal is $D(n = 1) \equiv D_1 = \ln 5 / \ln 3 = 1.4649$. A lot of other fractals (Koch, Vicsek [7, 12]) are characterized by the similar dimension $D_1$ independent on $n$, but their topology does not reflect specifics of radial wave expansion. The fact is that the dimension $D$ of a set of the anisotropic fractal hierarchy depends on $n$. Only at the correct choice of $\Delta R(n)$ in a selected direction, for example, by use of Eq.(3), we should expect $D(\Delta R(n)) = D_1$ for a wave process. Let us note that using of the suggested model of the anisotropic fractal is more suitable for solving the problem on the base of information-entropic analysis: $X_{i+1}(R)$ is characterized by its maximal value in case of minimal number of elements.

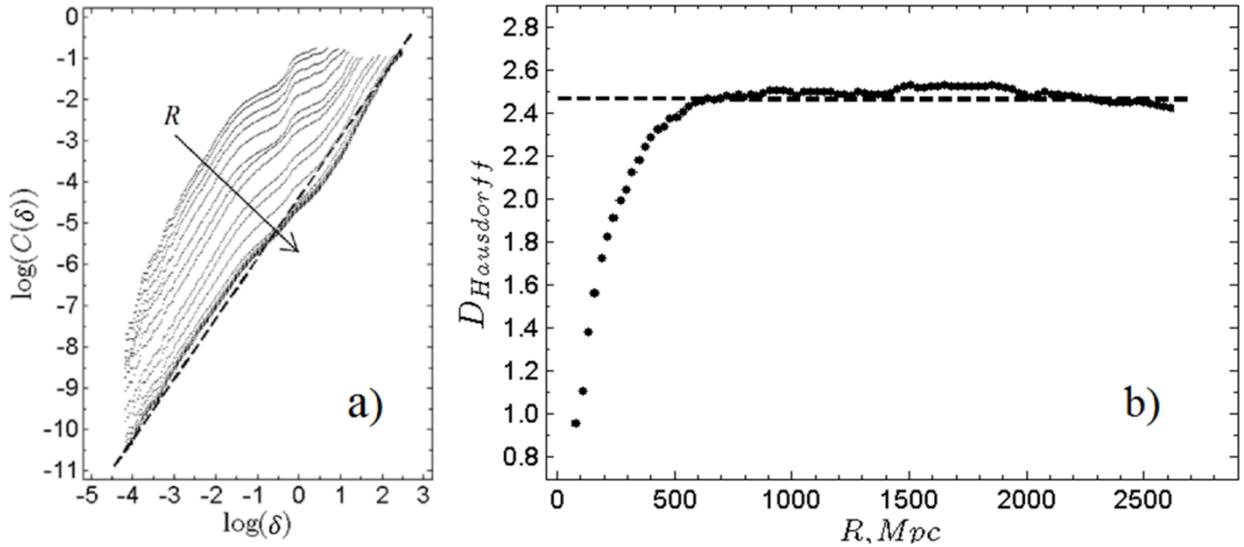

Fig. 4. – a) Dependence of correlation integral on scale of measurement according to [13]. Results have been approximated by the dashed line $\log(C(\delta)) = D * \log(\delta)$, where $D$ is equal to the theoretically defined dimension $D_1 = 0.4649$, $0 < R \leq 25\ Mpc$.
b) $D$ is the Hausdorff dimension of galaxies with a huge number of elements $N = 1.8 \cdot 10^6$ [14]. The spatial step is given in the interval $10^{-6} \div 10^{-3}$. Coordinates of points have been calculated for $\Delta R = 28,3\ Mpc$ according to Eq.(3) at $\Delta H = 5.5\%$.

Indeed, dimension $D_1$ can be established only for a certain spatial scale (Fig. 4a). Fig. 4b shows that the suggested model can be also used for the description of fractal distribution of galaxies in a space with zero-gravity regions. A set of galaxies forms a fractal surface ($d = 2$). Asymptotical scaling factor corresponds to its theoretical value $\gamma_* = D - d = 0.4649$.

Near the Local Group ($1 \leq R \leq 3\ Mpc$) values of recession velocity are insignificant, so, $\gamma < \gamma_*$ (Fig. 4a). Relative velocity of galaxies $\frac{dX_{i+1}}{dR} = \frac{dX_{i+1}/dt}{dR/dt}$ in astrophysical observations can be either positive or negative [15]. The global expansion of the Universe corresponds to $\gamma_* = 0.4649$. Fig. 5 shows the relative velocity-distance diagram for a set of values $10 \leq R_* \leq 200$ (in units of Mpc). We have accepted $\Delta H = 5\%$ and obtained $H = 74.9\ Mpc^{-1}$, where $\Delta R = 2\pi/\Delta H = 1.256\%$.

Near the Local group of galaxies (small values of $R_*$), values of scaling factor are also small ($\gamma < \gamma_*$). Clusters of galaxies are characterized by a set of values of $R_*$ and asymptotic behavior $\gamma \to \gamma_*$. Maximal radial velocity of galaxies is determined by the global value of the Hubble constant $H_0$. Let us note that fractal dimension of the set $X_{i+1}$ calculated by use of Eq.(2) is maximal at $\gamma = \gamma_*$ in accordance with Eq.(3).

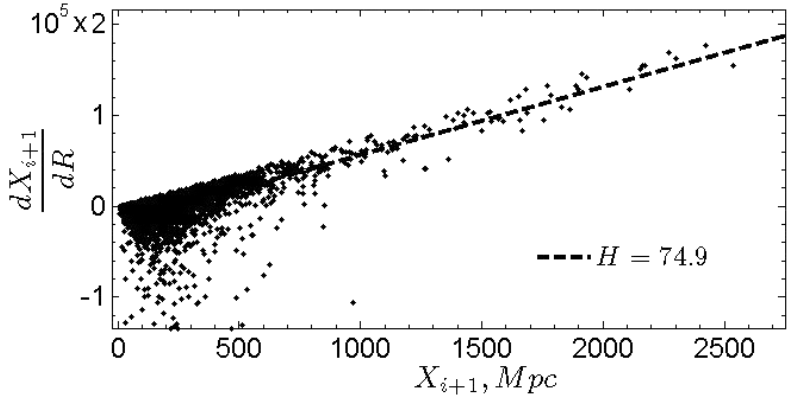

Fig. 5 – Existence of the global Hubble constant (according to our theory).

In our spatial model, values of acceleration can be defined as $H = H(R_*)$. Values of $H(R)$ determined with different accuracy are shown in Fig. 6. To compare of our data with theoretical results, we have accepted $R_* \sim R$ because only in this case we can state difference between values of distance and corresponding coordinate. The dependence $H = H(R_*)$ is characterized by saturation in agreement with the already noted asymptotic behavior to the global value $H_0$ with the given accuracy.

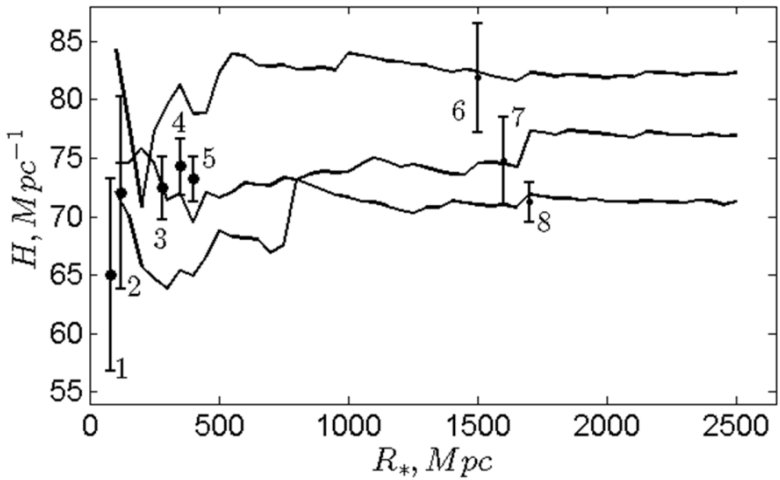

Fig. 6 – Theoretical and observed values of the Hubble constant: 1 – [16], 2 – [17], 3 – [18], 4 – [19], 5 – [10], curves: 6 – ($\Delta H = 5,5\%$), 7 – ($\Delta H = 5\%$), 8 – ($\Delta H = 2,4\%$) – our results. Theoretically, in case of $\Delta H = 2,4\%$ we accept $\Delta R = \pi/\Delta H = 1.3\%$ for fulfillment the condition $\Delta R < \Delta H$.

Using $H(1/c) = V_0 \cdot H(1/m)$ we can analyze time evolution of processes in 4-dimensional Minkowski space via characteristic time $1/H \approx 4 \cdot 10^{17} c$, however, in cosmology, observations of changing of space are more available. In analogy with Eq.(3), we can use the relation between accuracy of measurements of time and frequency. The approach used for determination of $H_0$ can be also applied for analyzing of processes with high density of matter and small values of relaxation time. In this case, a value which is an analogue of the threshold parameter $R_*$ should be chosen considering not only gravitational interactions, but another (for example, strong) types of interactions should be also taken into account.

**Main conclusions.** Galaxies move chaotically around galactic formations with zero gravitation and form fractal sets. Description of formation of clusters (self-organization) and their possible fractal structure is the fundamental result of physics of open systems. Fractal dimension of a set of galaxies increases at growth of size of this set (consisting of groups and clusters of galaxies). Limit value of fractal dimension determined theoretically corresponds to data describing structure of millions of galaxies.

Velocity of galaxies relative to a homogeneous isotropic medium with a wave structure can be either positive or negative in groups with small fractal dimensions tending to their limit value which is only positive in the chosen direction from the observer.

Taking into account a wave character of distribution of galaxies let us define a relationship between accuracy of determination the Hubble constant and coordinates of a certain galaxy. On the base of this analysis we can describe the accelerated recession of galaxies within global scale of the Universe fixed in cosmological observations. Further growth of the spatial scale and related with this growth phenomenon of relativistic redshift with zero frequency of radiation lead to an unobservable picture of the global Universe. Expansion of the Universe is a chaotic and ergodic process, therefore, time evolution of the Universe repeats laws of its spatial evolution. The problem of infinite forces arising at description of distribution of matter with nowhere vanishing average density in Newton's gravitation theory can be solved by considering clusters of galaxies separated by a zero gravity space. The discrete equations of fractal evolution of measure used in this theory do not contain singularities that are typical for metric space models by Einstein. The metric of the Universe can be described by use of fractal geometry which is a special branch of the non-Euclidean geometry.

References
1. Landau L. D. (ed.). The classical theory of fields. Vol. 2. *Elsevier* (2013).
2. Luminet J. P. *et al.* Dodecahedral space topology as an explanation for weak wide-angle temperature correlations in the cosmic microwave background. *Nature* **425**, 593-595 (2003).
3. Chernin A. D. Dark energy in the nearby Universe: HST data, nonlinear theory, and computer simulations. *Physics-Uspekhi* **56**(7), 704 (2013).
4. Byrd G. *et al.* Paths to dark energy: theory and observation. *Walter de Gruyter* (2012).
5. Chernin A. D. Dark energy and universal antigravitation. *Physics-Uspekhi* **51**(3), 253-282 (2008).
6. Arkani-Hamed N., Dimopoulos S., Dvali G. Phenomenology, astrophysics, and cosmology of theories with submillimeter dimensions and TeV scale quantum gravity. *Physical Review D* **59**(8), 086004 (1999).
7. Feder J. Fractals. *Springer Science & Business Media* (2013).
8. Peebles P. J. E. The large-scale structure of the universe. *Princeton university press* (1980).
9. Nicolis G., Prigogine I. Exploring complexity. *W.H. Freeman and Company/New York* (1989).
10. Riess A. G. *et al.* A 2.4% Determination of the local value of the Hubble constant based on observations with the NASA/ESA Hubble Space Telescope, obtained at the Space Telescope Science Institute, which is operated by AURA, Inc., under NASA contract NAS 5-26555. *The Astrophysical Journal* **826**(1), 56 (2016).
11. Zhanabaev Z. Z. Fractal model of turbulence in the jet. *Proceedings of the SB Acad.of Sci. USSR. Technical science series* **4**, 57-60 (1988). *in Russian*.
12. Metz V. How many diffusions exist on the Vicsek snowflake? *Acta Applicandae Mathematica* **32**(3), 227-241 (1993).
13. Karachentsev I. D. *et al.* A catalog of neighboring galaxies. *The Astronomical Journal* **127**(4), 2031 (2004).
14. Alam S. *et al.* The eleventh and twelfth data releases of the Sloan Digital Sky Survey: final data from SDSS-III. *The Astrophysical Journal Supplement Series* **219**(1), 12 (2015).
15. Karachentsev I. D. *et al.* The Hubble flow around the local group. *Monthly Notices of the Royal Astronomical Society* **393**(4), 1265-1274 (2009).
16. Jones M. E. *et al.* H0 from an orientation-unbiased sample of Sunyaev–Zel'dovich and X-ray clusters. *Monthly Notices of the Royal Astronomical Society* **357**(2), 518-526 (2005).
17. Freedman W. L. *et al.* Final results from the Hubble Space Telescope key project to measure the Hubble constant. *The Astrophysical Journal* **553**(1), 47 (2001).
18. Efstathiou G. H0 revisited. *Monthly Notices of the Royal Astronomical Society* **440**(2), 1138-1152 (2014).
19. Freedman W. L. *et al.* Carnegie Hubble program: a mid-infrared calibration of the Hubble constant. *The Astrophysical Journal* **758**(1), 24 (2012).